\pdfoutput=1
\documentclass[fleqn,usenatbib,usedcolumn]{mnras}

\usepackage{amssymb} 
\usepackage{times} 
\usepackage{aas_macros} 
\usepackage[dvipsnames,usenames]{color}

\usepackage{longtable}
\usepackage{array}

\usepackage{pifont} 
\usepackage{rotating}
\usepackage[toc,page]{appendix}
\usepackage[dvipsnames,usenames]{color}

\newcommand{\kepler}{\textit{Kepler}}
\newcommand{\corot}{\textit{CoRoT}}

\newcommand{\msol}{M$_\odot$}
\newcommand{\rsol}{R$_\odot$}
\newcommand{\lsol}{L$_\odot$}
\newcommand*{\Resize}[2]{\resizebox{#1}{!}{$#2$}}%

\def\muHz{\mu{\rm Hz}}
\def\note #1]{\noindent{\bf #1]}}

\def\sgnk2{{\rm sgn\left(K^2\right)}}
\def\rmd{{\rm d}}

\title[Helium glitches in red-giant stars]{On using dipolar modes to constrain the helium glitch in red-giant stars} 
\author[Dr\'eau et al.]{G. Dr\'eau$^{1,6}$, M. S. Cunha$^{2,3}$, M. Vrard$^{2,4}$, P. P. Avelino$^{2,3,5}$\\
$^1$ Magistère de Physique Fondamentale, Université Paris-Saclay, Bât. 625 91405 Orsay CEDEX, France\\
$^{2}$Instituto de Astrof\'\i sica e Ci\^encias do Espa\c co, Universidade do
Porto, CAUP, Rua das Estrelas, PT4150-762 Porto, Portugal\\
$^{3}$ School of Physics and Astronomy, University of Birmingham, Birmingham, B15 2TT, United Kingdom\\
$^{4}$Department of Astronomy, The Ohio State University, 140 West 18th Avenue, Columbus OH 43210, USA\\
$^{5}$ Departamento de Física e Astronomia, Faculdade de Ciências, Universidade do Porto, Rua do Campo Alegre 687, PT4169-007 Porto, Portugal\\
$^{6}$LESIA, Observatoire de Paris, PSL Research University, CNRS, Universit\'e Pierre et Marie Curie,
 Universit\'e Paris Diderot,  92195 Meudon, France\\
}

\date{Accepted XXX. Received YYY; in original form ZZZ}

\pubyear{2020}

\begin{document}
	\label{firstpage}
	\pagerange{\pageref{firstpage}--\pageref{lastpage}}
	\maketitle
\begin{abstract}
The space-borne missions \corot\ and \kepler\ have revealed numerous mixed modes in red-giant stars. These modes carry a wealth of information about red-giant cores, but are of limited use when constraining rapid structural variations in their envelopes. This limitation can be circumvented if we have access to the frequencies of the pure acoustic dipolar modes in red giants, {\it i.e.} the dipole modes that would exist in the absence of coupling between gravity and acoustic waves. We present a pilot study aimed at evaluating the implications of using these pure acoustic mode frequencies in seismic studies of the helium structural variation in red giants. The study is based on artificial seismic data for a red-giant-branch stellar model, bracketing seven acoustic dipole radial orders around $\nu_{\rm max}$. The pure acoustic dipole-mode frequencies are derived from a fit to the mixed-mode period spacings and then used to compute the pure acoustic dipole-mode second differences. The pure acoustic dipole-mode second differences inferred through this procedure follow the same oscillatory function as the radial-modes second differences. The additional constraints brought by the dipolar modes allow us to adopt a more complete description of the glitch signature when performing the fit to the second differences. The amplitude of the glitch retrieved from this fit is 15$\%$ smaller than that from the fit based on the radial modes alone. Also, we find that thanks to the additional constraints, a bias in the inferred glitch location, found when adopting the simpler description of the glitch, is avoided.
\end{abstract}

\begin{keywords}
	stars: evolution -- stars: interiors -- stars: oscillations
\end{keywords}

\section{Introduction\label{introduction}}

 {The space-borne missions $\corot$ \citep{baglin06} and $\kepler$ \citep{gillilandetal10} opened a new window of opportunities to enhance our understanding of stellar physics by providing unrivalled high-quality data. Thanks to the high precision of space-based photometric time-series, asteroseismology has become an increasing powerful tool to probe the internal structure and dynamics of stars. A common way to explore the constraining power of the seismic data is to include the frequencies of the oscillation modes or specific combinations of these, in the set of observations used in model-data comparisons \citep[e.g.][]{cunhaetal07,lebreton14,reese16}}.   {However, this procedure suffers from specific degeneracies that can limit the precision with which stellar properties are inferred. In particular, the fact that the impact on stellar observables from changing the stellar mass can be mimicked by changes in the initial helium abundance can severely hamper the determination of the stellar mass when the fractional helium abundance is not well constrained, as discussed in the context of different types of pulsating stars \citep[e.g.][]{cunha03,baudin12,lebreton14}.}  {Therefore, having a precise determination of the helium abundance can help to improve the precision with which the stellar properties are inferred. In principle, the helium content of the envelope can be estimated from the seismic signature of the glitch associated with the helium ionization region \citep{gough90}. Glitches are sharp variations in the structure of a star, taking place on a scale comparable or smaller than the local wavelength of the oscillation, and they introduce a modulation in the observed oscillation frequencies - the glitch seismic signature. This signature can be used as a diagnostic of the region where the glitch occurs, containing information on its location, as well as on how significantly and sharply the structure varies at that location. The signature from the helium glitch has been identified in  the  oscillation frequencies of the Sun \citep{monteiro05,Houdek07},  main-sequence stars \citep{mazumdar14,verma14,verma19}, and red giants \citep{miglio10,broomhall14, vrard15, corsaro15}. }

 {Solar-like oscillations are stochastically excited by near-surface convection. In a star with a spherical equilibrium (thus non-rotating and without a magnetic field), the oscillation frequencies $\nu_{n,\ell}$ are characterised by the radial order $n$ and the degree $\ell$. In main-sequence solar-like pulsators, the observed modes have an acoustic nature and their frequencies are approximately equally spaced in the power spectrum \citep{aerts10}. However, in red-giant stars gravity waves trapped in the stellar core are coupled with pressure waves trapped in the stellar envelope through the small evanescent region that separates the two cavities \citep{dupret09}. This coupling is possible when the underlying perturbation is non-radial and gives rise to mixed modes. Although mixed modes provide valuable information about the stellar core, the information they carry on the stellar envelope is more difficult to access than when the modes are purely acoustic. This, in turn, limits our ability to infer information about the helium glitch located in the envelopes of red-giant stars. \\}

 {In the specific case of the helium glitch, the structural variation is best seen in the first adiabatic exponent, $\gamma_{1}$, defined by}
\begin{equation}
    \gamma_{1} = \left(\frac{\mathrm{d}\log P}{\mathrm{d}\log \rho}\right)_{s},
\end{equation}
 {where $P$ and $\rho$ are pressure and density, respectively, and the subscript $s$ indicates that the derivative is taken at constant entropy. The value of $\gamma_1$ changes rapidly with depth at the location of the helium second ionization, inducing a rapid variation in the adiabatic sound speed.  It has been commonly assumed that the helium glitch signature  arises from the dip in $\gamma_{1}$ caused by the helium second ionization \citep{monteiro05, Houdek07}. However, based on a detailed study of the glitch signatures in main-sequence model frequencies,  \cite{verma14} have argued that the local maximum of $\gamma_{1}$, between the helium second and first ionization zones, provides a better representation of the location of the helium glitch in main-sequence stars. A similar conclusion was reached for red giants by \cite{broomhall14}.}

 {When the oscillations are purely acoustic, such as in main-sequence solar-like pulsators, the helium glitch signature has a well understood oscillatory behaviour \citep[e.g.][]{Houdek07} that is often best captured by computing the second frequency differences \citep{gough90}, 
\begin{equation}
\Delta_{2} \nu_{n,\ell}^{\rm a} = \nu_{n+1,\ell}^{\rm a}-2\nu_{n,\ell}^{\rm a}+\nu_{n-1,\ell}^{\rm a},
\label{secdif}
\end{equation}
where we used the superscript `a' on the frequencies to emphasise that we are referring to pure acoustic modes.
However, as previously mentioned, in red-giant stars non-radial modes are mixed (only the radial modes are purely acoustic). Previous studies aiming at detecting and characterising the signatures of the helium glitch in red giants have thus been based on radial modes alone \citep{miglio10,vrard15,corsaro15}. This limits significantly the number of frequencies available to constrain the properties of the glitch. }

In this work we present the results of a pilot project aimed at understanding whether pure acoustic dipolar frequencies retrieved from fitting the period spacing may strengthen the  constraints on the helium glitch's properties in red-giant stars. Moreover, we want to understand if using the radial modes alone to characterise the glitch properties, as done in previous works, may introduce biases in the results.  The study is based on model simulated data. In Sec.~\ref{data} we describe the model adopted for our data simulation, the frequency range of the modes, and the uncertainties considered on the simulated frequencies. In Sec.~\ref{dipole} we present the method used to retrieve the pure acoustic frequencies from the simulated period spacing. In Sec.~\ref{fits} we compare the helium glitch properties inferred from the second differences using different analytical expressions and different sets of modes, highlighting the improvement brought by the inclusion of the pure acoustic dipolar frequencies. In Sec.~\ref{discussion} we discuss the results and identify further tests to be carried out in future work.

\section{Simulated red-giant pulsation data}
\label{data}
Based on a study of a series of  red-giant-branch (RGB) stellar models, \cite{broomhall14} concluded that for models with $\nu_{\rm max} > 70\mu$Hz the $\ell=1$ mixed modes are not useful to constrain the properties of the helium glitch, since even the modes of lowest inertia have their frequencies significantly perturbed compared to those of pure acoustic modes. Our aim is to infer the pure acoustic frequencies for the $\ell=1$ modes, and to use them to constrain the properties of the helium glitch.  Therefore, we chose to simulate pulsation data for an RGB model with $\nu_{\rm max}\sim 105$~$\muHz$, well in the $\nu_{\rm max}$ range where the frequencies of the mixed $\ell=1$ modes cannot be used to achieve that goal. To that effect, we used one of the models studied by \cite{cunha19}, whose global properties are shown in Table~\ref{tab_model}. This model has been extracted from an evolution series of 1~$M_\odot$ models computed with the evolution code ASTEC \citep{jcd08a}. The corresponding pulsation frequencies have been derived with the adiabatic pulsation code ADIPLS \citep{jcd08b}. 

According to \cite{mosser12}, the underlying power distribution of the observed modes in an RGB star can be described by a Gaussian centred on $\nu_{\rm max}$ with a full width at half maximum of
\begin{equation}
\delta \nu_{\rm env} = 0.66\nu_{\rm max}^{0.88}.
\end{equation}
To keep a realistic approach, in our simulations we shall consider only a limited number of radial orders within this power envelope. Based on the stars analysed by \cite{corsaro15}, we chose to consider 8 radial modes distributed around $\nu_{\rm max}$, which allow us to compute six radial-mode second differences. In the case of our model, that represents modes within an envelope of 1.7$\delta \nu_{\rm env}$, slightly larger than the 1.5$\delta \nu_{\rm env}$ considered in the theoretical work by \cite{broomhall14}. In addition, we consider the dipole mixed-modes with frequencies within the same range,  which brackets 7 pure acoustic $\ell=1$ frequencies that will need to be inferred.

Based on the radial and dipolar mode frequencies for the adopted red-giant model, we generate sets of artificial data. To that end, we perturb the mode frequencies by considering a normal distribution for the errors on the frequencies with zero average and two possible standard deviations, namely: (1) $\sigma = 0.005$~$\mu$Hz, which will be referred to as the best-case scenario and (2) $\sigma = 0.01$~$\mu$Hz, which corresponds to a more realistic scenario, considering the results from the analysis of stars observed by Kepler for a period of 4~yrs  {\citep[Table~(B.1-B.49); Table~(1) and (A.1)]{corsaro15,vrard18}.}

\begin{table}
	\caption{Properties of the stellar model considered in this work. The frequency of maximum power is $\nu_{\rm max}$=105~$\mu$Hz.
	}
	\begin{minipage}{0.48\textwidth}
		\resizebox{\linewidth}{!}{%
			\begin{tabular}{|l|c|c|c|c|c|}
				\hline
				{\bf Model} & {\bf Mass}  & {\bf Radius}  & {\bf Effective Temp.} & {\bf Luminosity}  & {\bf Age}  \\
					 & \msol &  \rsol &  K & \lsol & Gyrs \\			
				\hline
				RGB &1.0& 5.8 & 4624 & 14.0  & 11.37 \\
				\hline
			\end{tabular}
		}
	\end{minipage}
	\label{tab_model}
\end{table}

\section{Inferring the ``pure" acoustic dipole frequencies}
\label{dipole}
\label{pure}
Gravity modes of a given degree $\ell$ in a non-rotating star have asymptotically equally spaced periods whose difference is given by the asymptotic period spacing \citep{tassoul80},
\begin{equation}
\Delta P_{\rm as}= \frac{2\pi^2}{\omega_{\rm g}},
\label{psasymp}
\end{equation}
where,
\begin{equation}
\omega_{\rm g}\equiv\int_{r_1}^{r_2}\frac{LN}{r}\rmd r,
\label{omegag}
\end{equation}
$N$ is the buoyancy frequency, $L^2=\ell(\ell+1)$, and $r_1$ and $r_2$ are the radii at the inner and outer turning
points that define the propagation cavity of the g mode,  respectively. 

In red-giant stars, gravity modes couple with acoustic modes and give rise to mixed modes.  At constant degree $\ell$, mixed modes are not equally spaced in period, as would be expected in purely gravity-mode pulsators \citep{tassoul80}. Instead, the period spacing shows rapid variations at the frequencies where the pure acoustic modes would appear if no coupling existed \citep[e.g.,][]{mosser12}. These pure acoustic frequencies are approximately equally spaced in frequency by the asymptotic large frequency separation.
To infer the pure acoustic frequencies of the dipolar modes, we consider the analytical expression derived by \cite{cunha15} to describe the mixed-mode period spacing, $\Delta P$, defined as the difference between the periods of consecutive mixed modes of the same degree $\ell$. Adopting the formulation presented in \cite{cunha19}, the period spacing for dipole modes  ($\ell =1$) in a non-rotating red-giant star without core glitches is
\begin{eqnarray}
\label{ps_coupling_3}
{{\frac{\Delta P}{\Delta P_{\rm as}}\Resize{7.5cm}{\approx
			\hspace{-0.08cm}\left[1+{
				{\frac{\omega^2}{\omega_{\rm g}}\frac{q}{\omega_{\rm p}}\left[\sin^2\left(\frac{\omega-\omega_{n,1}^{\rm a}}{\omega_{\rm p}}\right)+q^2\cos^2\left(\frac{\omega-\omega_{n,1}^{\rm a}}{\omega_{\rm p}}\right)\right]^{-1}}}+Q\left(\omega\right)\right]^{-1}}}}
\hspace{-0.2cm},\hspace{-0.1cm}
\end{eqnarray}
where the function $Q(\omega)$ is given by
\begin{equation}
Q(\omega)=\frac{q_1\alpha\omega^2}{2\pi\nu_{\rm max}\omega_{\rm g}}\left[q^2\cot\left(\frac{\omega-\omega_{n,1}^{\rm a}}{\omega_{\rm p}}\right)+\tan\left(\frac{\omega-\omega_{n,1}^{\rm a}}{\omega_{\rm p}}\right)\right]^{-1}.
\end{equation}
In the expressions above , $\omega=2\pi\nu$ is the angular frequency and 
 \begin{equation}
\omega_{\rm p}=\left(\int_{r_3}^{r_4}c^{-1}\rmd r\right)^{-1},
\label{omega_p}
\end{equation} 
where, $r_3$ and $r_4$ are the turning points of the p-mode cavity. This quantity is approximately equal to twice the  asymptotic large frequency separation (and never smaller than that). Moreover, $q$ is the coupling factor which is allowed to depend linearly on frequency ({\it cf.} \cite{cunha19}) through
  \begin{equation}
 q=q_1\left[\alpha\left(\frac{\omega}{2\pi\nu_{\rm max}}-1\right)+1\right],
 \end{equation}
 where $\nu_{\rm max}$ is the cyclic frequency at maximum power \citep[e.g.][]{chaplin13}. Thus, $q_1$ represents the coupling factor at the maximum oscillation power, while $\alpha$ determines how strongly $q$ depends on frequency. Finally, the frequencies $\omega_{n,1}^{\rm a}$ are the set of pure acoustic frequencies of radial orders $n$ that shall be estimated through the fitting of Eq.~(\ref{ps_coupling_3}) to the model simulated data.

\begin{figure} 
	\begin{minipage}{0.97\linewidth}  
		\rotatebox{0}{\includegraphics[width=1.0\linewidth]{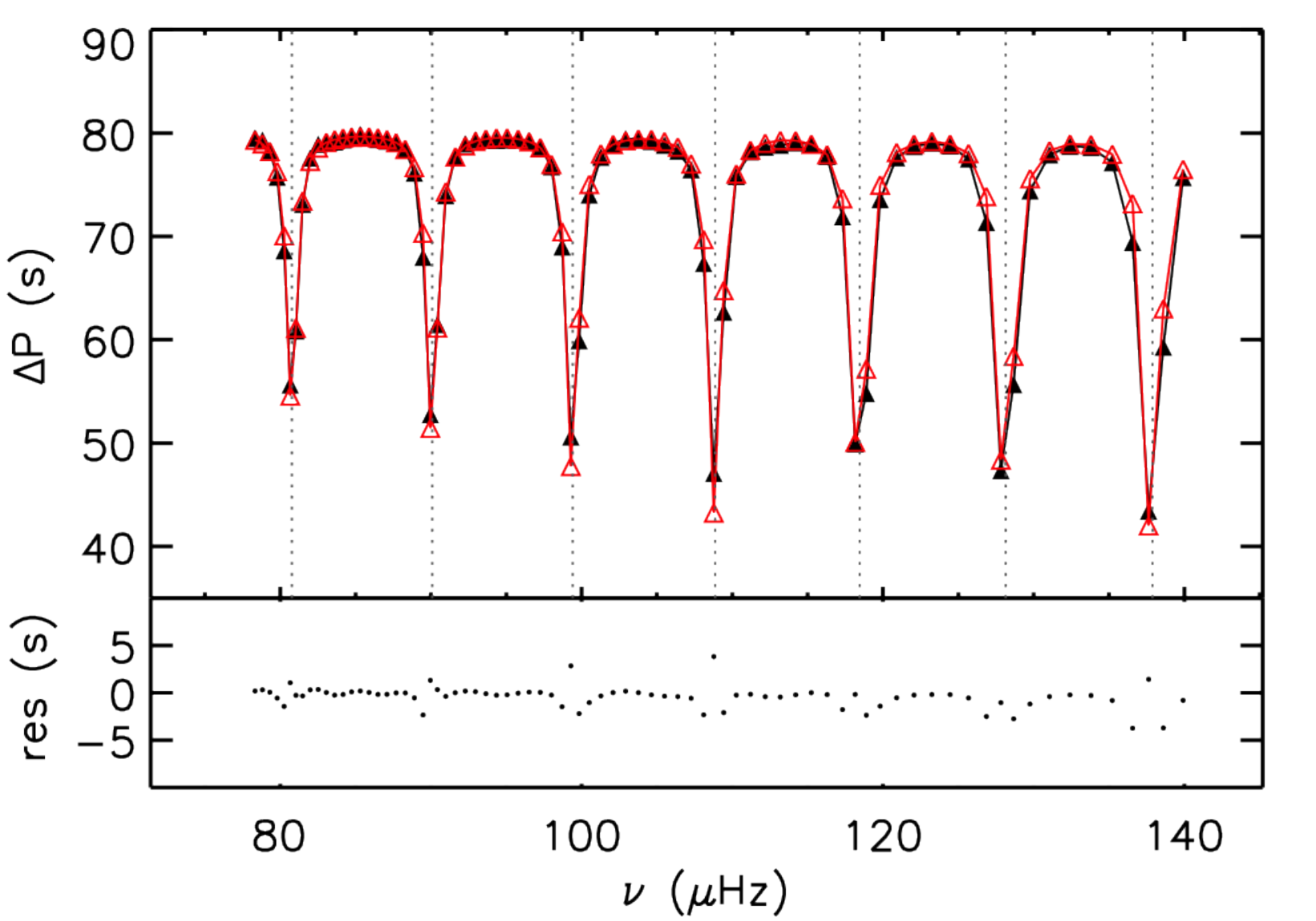}}
	\end{minipage}
	\caption{Top: Comparison between the period spacing derived from the frequencies returned by the pulsation code ADIPLS (filled triangles, black) and those obtained from  Eq.~(\ref{ps_coupling_3}) with the parameters of the best-fit model (open triangles, red). The dotted vertical lines show the position of the frequencies of the pure acoustic modes for the best-fit model. Bottom: the residuals computed as  (`ADIPLS period spacing'$-$`analytical period spacing').  {The frequencies used in the fit lie in the range [74.98, 141.54] $\muHz$.}
	}
	\label{dp}
\end{figure}

Figure~\ref{dp} shows the result from the fit of the analytical expression given by Eq.~(\ref{ps_coupling_3}) to the unperturbed model data. The fit was performed using the
python module {\it emcee} implementation of the affine-invariant ensemble
sampler for Markov chain Monte Carlo \citep{emcee} with the likelihood defined by,
\begin{equation}
{\mathcal L} = \frac{1}{\left({2\pi}\sigma_{\rm fit}^2\right)^{N/2}}\exp\left(-\frac12 \chi^2\right),
\end{equation}
where the uncertainty $\sigma_{\rm fit}$ was left as a free parameter and
\begin{equation}
\chi^2=\sum_i\left(\frac{\Delta P_i-\Delta P_{{\rm\small ADIPLS},i}}{\sigma_{\rm fit}}\right)^2,
\end{equation}
with the subscript ``ADIPLS" indicating the  period spacing derived from the frequencies returned by ADIPLS.

\section{Inferring the properties of the glitch}
\label{fits}
\subsection{Glitch properties in the error-free analysis}
According to the analysis by \cite{Houdek07}, the glitch properties can be inferred from the second differences through the fit of the frequency dependent function,
\begin{equation}
\Delta_{2}\nu = A\,\nu\,{\rm e}^{-2\Delta_{\rm II}^{2}\left(2\pi\nu\right)^2}\cos{\left[2\,\left(2\pi\tau\nu+\epsilon\right)\,\right]}+\mathcal{F(\nu)}
\label{fit}
\end{equation}
where $A$ and $\Delta_{\rm II}$ are related to the glitch amplitude and width, respectively. Here, $\epsilon$ is a phase parameter and $\tau$ is the acoustic depth of the glitch defined by, 
\begin{equation}
     {\tau = \int_{r_g}^{R}{\frac{dr}{c}}}
\end{equation}
 {where $R$ is the seismic radius of the star \cite[see][for a discussion]{gough90} and $r_g$ is the radius at which the glitch is located.} Moreover, $\mathcal{F}$ is a smooth function of frequency that incorporates  effects from the near surface, the hydrogen ionisation zone and the wave refraction in the stellar core. Polynomial expansions in either $1/\nu$ or $\nu$ are usually adopted for $\mathcal{F}$ \citep[e.g.][]{Houdek07,verma14} when modelling acoustic glitches in main-sequence stars. However, given the limited number of radial modes observed in RGB stars, studies of the helium glitch in these evolved stars have so far considered only the first term in the expansion, assuming that the function $\mathcal{F}$ is a constant offset, $B$. In that way, these works constrained the 5 parameters $\{A,\Delta_{\rm II},\tau,\epsilon,B\}$,  with  typically 6 radial-mode second differences. The extraction of the frequencies of the purely acoustic dipole modes, if successful, would allow for the consideration of additional terms in that expansion. We shall discuss the impact on the inferred glitch parameters of the addition of such acoustic dipole-mode second differences to the fit. 

\begin{figure} 
	\begin{minipage}{1.\linewidth}  
			\rotatebox{0}{\includegraphics[width=1.0\linewidth]{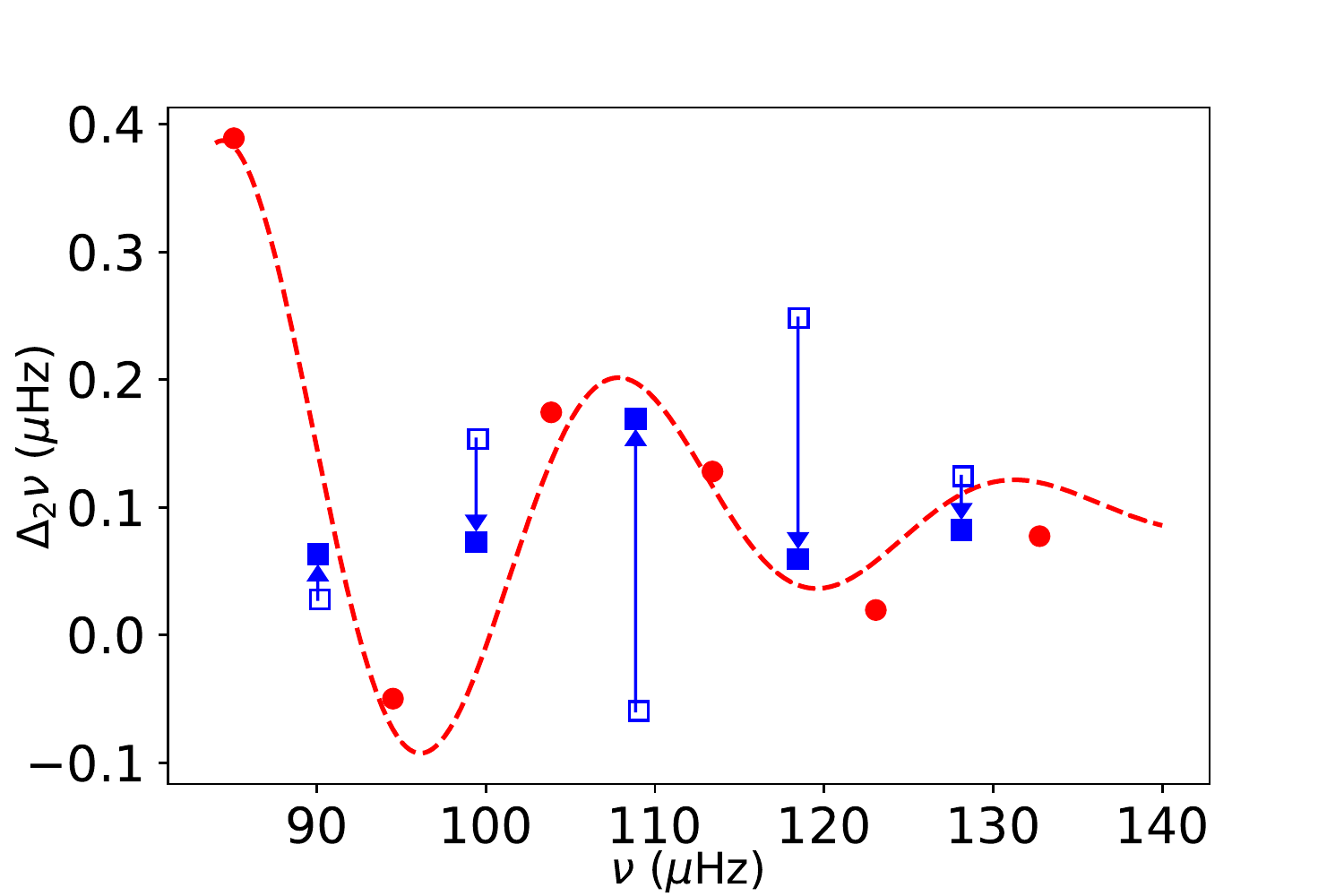}}
	\end{minipage}
	\caption{The second differences computed from the radial-mode frequencies (red, filled circles), from the purely acoustic dipole-mode frequencies (blue, filled squares) and from the frequencies of the dipole mixed modes with lowest inertia (blue, open squares). The arrows connect the two sets of dipole-mode second differences. The red-dotted curve shows the function described by Eq.~(\ref{fit}) with $\mathcal{F}$ taken as a constant, for the parameters producing the most likely fit to the radial-mode second differences only.
	}
	\label{sec_dif_1}
\end{figure}

 In Fig.~\ref{sec_dif_1} the second differences computed from the  $\ell=0$ mode frequencies of our model are shown along with those computed from the purely acoustic $\ell=1$ modes extracted in Sec.~\ref{pure}. For comparison, the second differences computed from the dipole mixed modes of lowest inertia are also shown (open symbols). Arrows connect the two sets of $\ell=1$ second differences. It is clear from simple eye inspection that the second differences computed from the purely acoustic dipole frequencies follow closely the oscillatory signature seen in the second differences computed from the radial modes, unlike the second differences computed from the mixed modes. 

\begin{figure*} 
	\begin{minipage}{1.0\linewidth}  
		\rotatebox{0}{\includegraphics[width=0.5\linewidth]{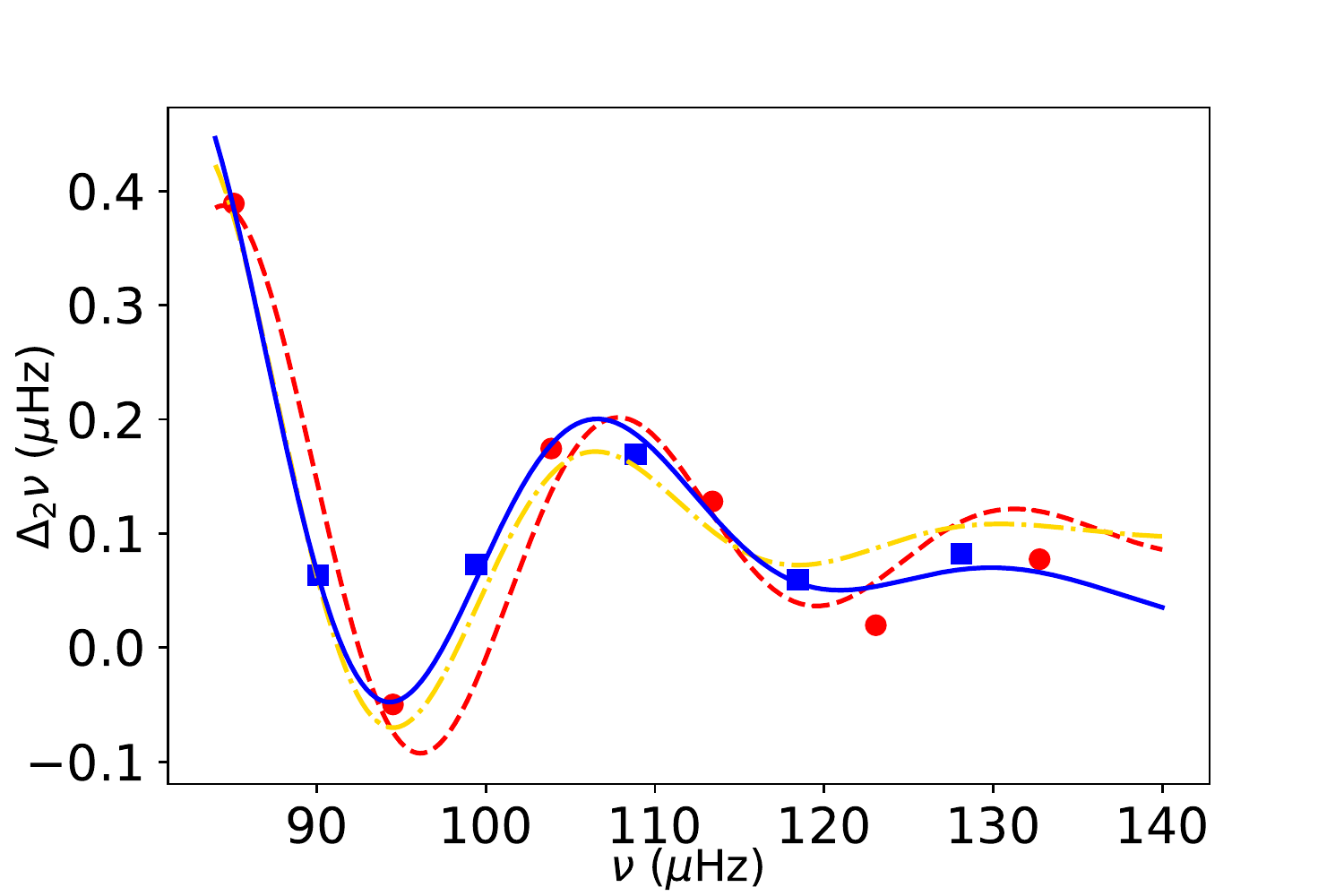}}
	\rotatebox{0}{\includegraphics[width=0.5\linewidth]{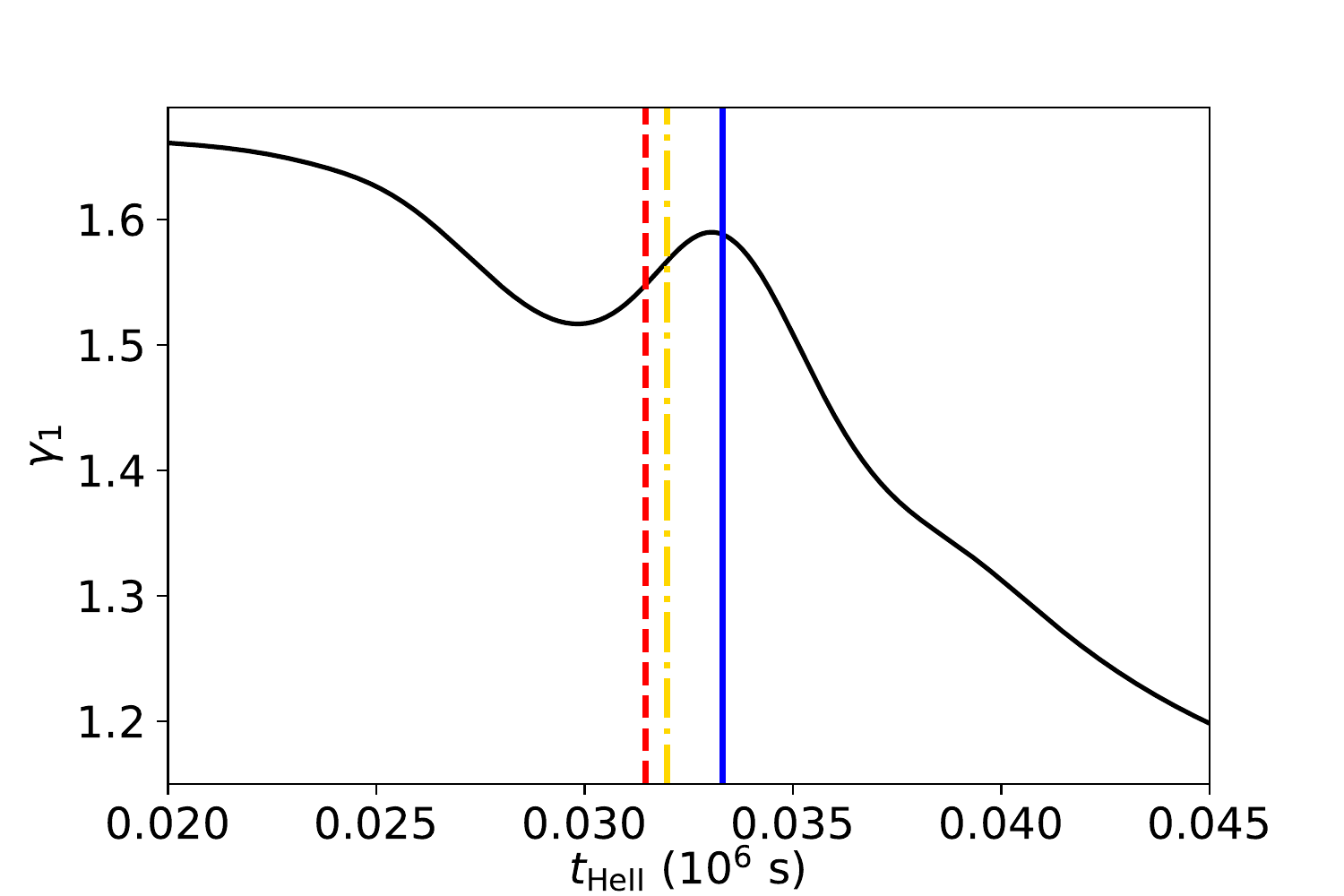}}
\end{minipage}
	\caption{Left: Best fits of Eq.~(\ref{fit}) to the second differences (symbols as in Fig.~\ref{sec_dif_1}). Results are shown for three different cases, namely: fit to the second differences computed from radial and purely acoustic dipole modes for $\mathcal{F}=B+D\nu^2$ (blue, continuous curve) and for  $\mathcal{F}=B$ (yellow, dashed-dotted curve). The red-dashed curve is the same as in Fig.~\ref{sec_dif_1}. Right: The first adiabatic exponent around the region of helium ionisation, with the identification of the position of the helium glitch inferred from the fits shown on the left panel (vertical lines; colour and line style as in left panel).
	}
	\label{dif_exp}
\end{figure*}

To explore the impact on the inferred glitch location of adding the pure acoustic dipole modes to the fit, we compare the results from fitting  Eq.~(\ref{fit}) to the second differences computed from the radial modes alone with those obtained from fitting, in addition, the inferred purely acoustic dipole modes. For the latter case we consider two options for the smooth function $\mathcal{F}$, namely, one in which  $\mathcal{F}=B+D\nu^2$, where $B$ and $D$ are constant parameters, and another in which $\mathcal{F}=B$.  {We did not consider a term varying proportionally to $\nu$ in the first option because we verified that taking this linear term into consideration would not improve the quality of the fit. Since considering such a term would add another free parameter without improving the fit, we decided not to include it.}
 {To quantify the goodness of the fits, we compute the reduced $\chi^2$ for each case, defined by}

\begin{equation}
    \chi^2_R = \frac{1}{N - m} \sum_{i = 0}^{N}{{\left(\frac{\Delta_{2}\nu_{{\small{\rm ADIPLS}},i} - \Delta_{2}\nu_{i}}{\sigma_{i}}\right)^{2}}}
    \label{eq:chi2}
\end{equation}
 {where $N$ is the number of data points, $m$ is the number of free parameters in the model in Eq.~(\ref{fit}), and $\sigma_{i}$ are the uncertainties on the data points. Note that our data points are derived from the adiabatic pulsation code ADIPLS, thus, in the error-free case discussed in this subsection they do not have associated uncertainties. As our aim is solely to compare the goodness of the fits, we choose to set $\sigma_{i} = 1$ for all $i$.  The results from the fits are shown on the left panel of Fig.~\ref{dif_exp} and the values of $\chi^{2}_{R}$ obtained in each case are summarised in Table \ref{table:chi2}. They show that the inclusion of the second differences derived from the pure acoustic dipolar frequencies improves the fit, in particular when the quadratic term is also included in the function $\mathcal{F}$.   In addition, we calculate $\chi^{2}_{R}$ for the solution found when fitting the radial modes alone, now including the purely acoustic $\ell = 1$ modes. This allows us to assess how far this solution is from reproducing the dipolar-mode second differences which were not accounted for in the fit. We find $\chi^{2}_{R}$ = 0.0041, a value comparable to the value found when only the radial modes are considered in the $\chi^{2}_{R}$  computation and much larger than the value of $\chi^{2}_{R}$ obtained for the best case shown in Table~\ref{table:chi2}}


The right panel  of Fig.~\ref{dif_exp} shows a comparison of the glitch location inferred in the three cases illustrated on the left panel of the same figure. We present the glitch location in terms of its acoustic radius, $t_{\rm HeII}$\footnote{HeII indicates second helium ionization}, rather than acoustic depth, $\tau$, so that it can be directly compared with the results from the study of \cite{broomhall14}. That is achieved by making the transformation $t_{\rm HeII}=T-\tau$, where $T$ is the total acoustic radius of the star estimated by $\left(2\left<\Delta\nu\right>\right)^{-1}$.
Here, $\left<\Delta\nu\right>$  is the average large frequency separation which we compute by determining the slope of the linear fit to the $\ell=0$ mode frequencies expressed as a function of radial order. The acoustic radius computed in this way provides an  estimator of the location of the glitch that is less biased by the unknown exact position of the surface radius \citep{jcd95,ballot04}.

The results from the theoretical studies by \cite{broomhall14} and by \cite{verma14b} both indicate that the glitch position corresponds to the local maximum of $\gamma_1$.  Figure~\ref{dif_exp} shows that only when both radial and dipole modes are considered and the smooth function is allowed to vary with the square of the frequency, the expected position of the glitch is accurately recovered for this mode. That is quantified in Table~\ref{distance} where the relative distance of the glitch location from the local maximum in $\gamma_1$ is provided for each of the three cases shown in Fig.~\ref{dif_exp}.

{As the abundance of helium is expected to be directly related to the glitch amplitude, we  also consider the impact on the inferred amplitude from fitting different sets of observables or using different fitting expressions. Different proxies for the glitch amplitude are adopted by different authors. Here we follow \citet{verma19}, and use the average glitch-signature amplitude defined by
\begin{equation}
    \langle A \rangle = \frac{\int_{\nu_1}^{\nu_2}A\,\nu\,{\rm e}^{-2\Delta_{\rm II}^{2}\left(2\pi\nu\right)^2}{\rm d}\nu}{\int_{\nu_1}^{\nu_2}\nu{\rm d}\nu},
\end{equation}
where $\nu_1=85.09\ \mu$Hz and $\nu_2=132.74\ \mu$Hz are the smallest and largest radial-mode frequencies that limit the frequency range considered in the study.
The results are shown in Table~\ref{distance}. We find that the average amplitude varies by a maximum of 26$\%$ when the three different cases are considered. The difference between the average amplitudes from the fit with the most complete expression (blue line in Fig.~\ref{dif_exp}, left) and the fit to the radial-order second differences only (red line in the same figure) is  $15\%$.}

\begin{table}
	\caption{ Values of the $\chi^2_{R}$ computed with Eq.~(\ref{eq:chi2}), for the three cases discussed in the text.
	}
	\begin{minipage}{0.48\textwidth}
		\resizebox{\linewidth}{!}{%
			\begin{tabular}{|l|c|}
				\hline
				\hspace{2cm}{ Fitting conditions} & {{$\chi^{2}_{R}$}}   \\
				\hline
				  $\ell=0$, constant $\mathcal{F}$ & 0.0054 \\
				 \hline
				  $\ell=0$ and $\ell=1$, constant $\mathcal{F}$ & 0.0016\\
				 \hline
				 $\ell=0$ and $\ell=1$, quadratic $\mathcal{F}$ \hspace{3.cm} & 0.00050\\
				\hline
			\end{tabular}
		}
	\end{minipage}
	\label{table:chi2}
\end{table}



\begin{table}
	\caption{Average glitch-signature amplitude, $\langle A \rangle$, and glitch acoustic radius, $t_{\rm HeII}$, inferred from the fits to the second differences, for the three cases discussed in the text. Also shown is the relative distance of the glitch to the local maximum of $\gamma_1$, ${\rm RD} \,=\, (t_{\gamma_1,{\rm max}}-t_{\rm HeII})/T$, where $t_{\gamma_1, {\rm max}}$= 33069~s and $T=52576$~s. The last row concerns the results for the median values derived from the Monte Carlo Simulations discussed in Sec.~\ref{mcs}.
	}
	\begin{minipage}{0.48\textwidth}
		\resizebox{\linewidth}{!}{%
			\begin{tabular}{|l|c|c|c|}
				\hline
				\hspace{2cm}{Fitting conditions} & $\langle A \rangle$ & $t_{\rm HeII}$ (s) & RD   \\
				\hline
				  $\ell=0$, constant $\mathcal{F}$ & 0.122 & 31435 & 0.0311\\
				 \hline
				  $\ell=0$ and $\ell=1$, constant $\mathcal{F}$ & 0.094 & 31947 & 0.0213\\
				 \hline
				 $\ell=0$ and $\ell=1$, quadratic $\mathcal{F}$ & 0.105 & 33297 & -0.0043\\
				\hline
				$\ell=0$ and $\ell=1$, quadratic $\mathcal{F}$, $MCS$ & 0.111 & 33236 & -0.0032\\
				\hline
			\end{tabular}
		}
	\end{minipage}
	\label{distance}
\end{table}

\subsection{Impact of uncertainties on the glitch location}
\label{mcs}
\begin{figure*} 
	\begin{minipage}{1.0\linewidth}  
		\rotatebox{0}{\includegraphics[width=0.50\linewidth]{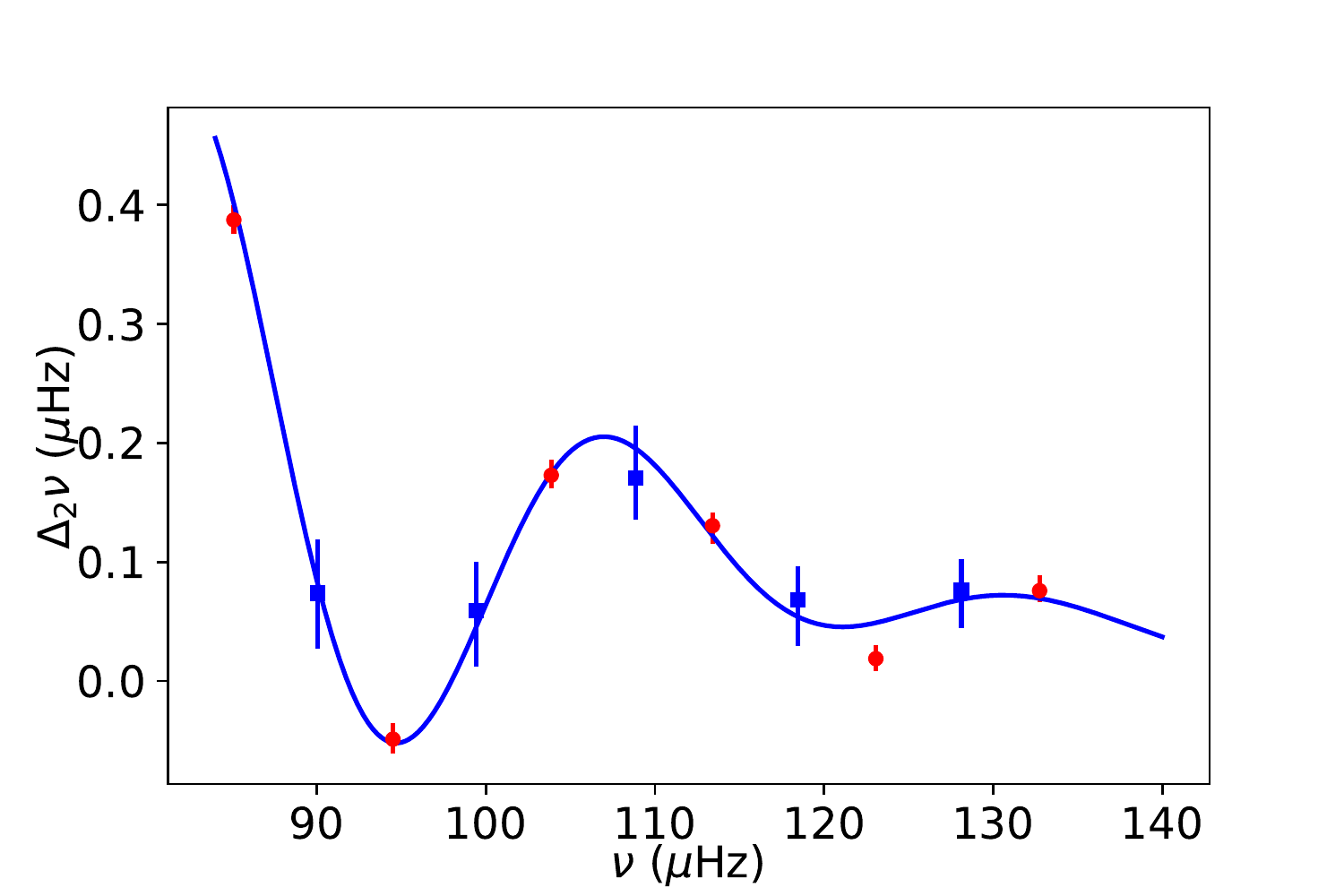}}
	\rotatebox{0}{\includegraphics[width=0.50\linewidth]{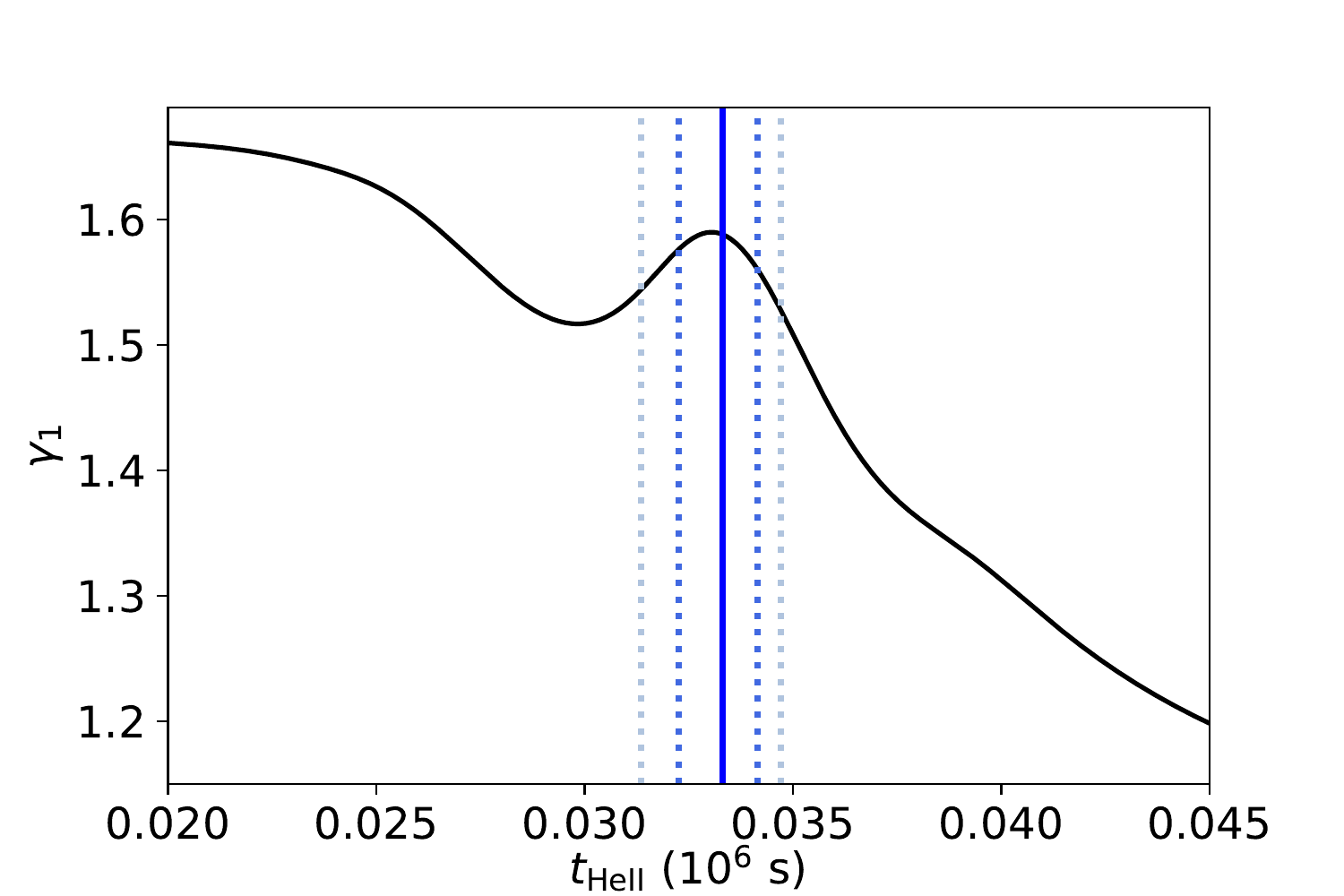}}
\end{minipage}
	\caption{Results from the Monte-Carlo simulations on the radial and mixed modes. Left: The blue curve represents the expression given by Eq.~(\ref{fit}) with $\mathcal{F}=B+D\nu^2$ where the parameters in the equation are taken to be the median values of the distributions resulting from the Monte-Carlo simulations in the best case scenario described in Sec.~\ref{data}. The red dots and the blue squares are the median values of the sets of second differences computed for radial and purely acoustic dipolar modes, respectively. The errorbars represent the 1~$\sigma$ interval around the median values. Right : The adiabatic exponent around the region of the helium ionisation. The vertical solid line correspond to the median value of the inferred position of the glitch and the dotted vertical lines show the $\pm$ 1$\sigma$ uncertainties around the median value, in darker blue for the best-case scenario and in lighter blue for the more realistic scenario described in Sec.~\ref{data}.
	}
	\label{fig:uncertainty_dif_fit}
\end{figure*}

So far we have fitted the model data without uncertainties to quantify the biases on the inferred glitch location introduced by the use of different sets of data and different fitting expressions. In reality the errors on the measured oscillation frequencies will propagate to the second differences. The errors in the second differences computed from the radial modes could be easily derived by linearly propagating the errors in the radial mode frequencies. However, in the case of the dipolar modes the situation is more complex, because the purely acoustic dipole modes are not directly observed, but rather inferred from the fit of Eq.~(\ref{ps_coupling_3}) to the mixed-mode period spacing. To properly account for this, we perform Monte Carlo simulations on the radial and the mixed modes, under the best-case and the more realistic scenarios described in Sec.~\ref{data}. For the radial modes we compute the second differences for each simulation directly from Eq.~(\ref{secdif}). In addition, we use the mixed modes  generated in each simulation to compute the corresponding period spacing which we then fit with Eq.~(\ref{ps_coupling_3}), leaving all parameters in this equation free to vary, to derive a simulated set of purely acoustic dipole modes. These purely acoustic modes are then used to compute the dipole second differences for each simulation.  We repeat the procedure  for a total of  {200} simulations and for each simulation we fit the resulting radial and dipolar second differences simultaneously with Eq.~(\ref{fit}), taking $\mathcal{F}=B+D\nu^2$. For a few simulations (7 out of the 200) the returned $\tau$ was larger than the Nyquist period and the fits were rejected. From the remaining fits, we obtain the probability distributions for the parameters in the fitted expression. 
 { The results are shown in Fig.~\ref{fig:uncertainty_dif_fit}. The uncertainties in the dipole-mode second differences are significantly larger than those in the radial-mode second differences (left panel), reflecting the additional error introduced by the fit of the analytical expression (Eq.~(\ref{ps_coupling_3})) to the period spacings. All but one second-difference values are within 1~$\sigma$ of the median curve, represented by Eq.~(\ref{fit}) with $\mathcal{F}=B+D\nu^2$ and the parameters taken to be the median values of the distributions resulting from the Monte Carlo simulations in the best case scenario. The exception is a radial-mode second difference found at 2.73 $\sigma$ away from the median curve.  
Moreover, when uncertainties are considered, the median value of the glitch acoustic radius inferred from the fitting of the combined radial-mode and dipole-mode second differences still lies near the local maximum of $\gamma_{1}$ (right panel), at a relative distance of -0.0032 (Table~\ref{distance}).}

\section{Discussion}
\label{discussion}



 {The analysis of space-based asteroseismic data has brought the study of acoustic glitches to a new standard. A number of recent works \citep{mazumdar12,mazumdar14,verma14,verma19} have shown that the properties of the glitch associated to the helium ionization zone can be successfully inferred in main-sequence stars through the fitting of the frequencies, or combinations of frequencies, from radial and non-radial, low-degree modes. However, in the case of red-giant stars the situation is more complex due to the mixed character of the non-radial modes. Since their frequencies deviate from those of pure acoustic modes they cannot reliably be used to infer the glitch parameters. The problem is more significant for stars with $\nu_{\rm max} \geq 70 \mu$Hz, as discussed by \cite{broomhall14}. One possible solution is to use only radial modes in the inference process. {However, as discussed in Section \ref{fits}, considering only the radial modes in the fit of the second differences limits the number of parameters allowed in the smooth component of the model, which, in turn, bias the inferred glitch parameters: the acoustic radius of the glitch $t_{\rm He_{II}}$ inferred from the fit deviates from the glitch expected location, at the local maximum in $\gamma_{1}$, and the average amplitude of the glitch signature can be affected (in the case of the present model, by about 15$\%$). Although the authors do not discuss it, inspection of the results from \cite{broomhall14} seems to indicate that the bias in the location inferred for the glitch is amplified towards more evolved red giants, with lower $\Delta\nu$ ({\it cf.} comparison of the two panels in their Figure 3).} }

 {In our study, we have tested a new procedure aimed at improving the characterisation of the glitch, which consists in first inferring what the frequencies of pure acoustic dipole modes in a red-giant star would be, if no mode coupling existed, and then using those frequencies to construct the coupling-free dipole second differences. To infer the pure acoustic dipole  frequencies, we followed the method proposed by \cite{cunha19}. We have shown that the second differences computed from the pure acoustic dipole mode frequencies inferred by this method follow the same oscillatory function as the second differences computed from the radial modes. The end product of this process is  an additional set of observational constraints that, in turn, allows us to adopt a more complete description for the smooth component $\mathcal{F}$ of the fitting function given in Eq.~(\ref{fit}). 
 Indeed, this study brings to light that the use of a constant smooth component in the fitting expression is not appropriate. A significant curvature in the smooth component is not entirely surprising because in red giants the observed frequencies are further away from the asymptotic limit than in main sequence stars. Thus, one of the main advantages of the addition of the dipolar modes is indirect, in that it allows us to fit a more complete expression (i.e., to give the necessary freedom for the smooth part to be properly accounted for and the actual glitch signature to be extracted). }
  With the addition of the pure acoustic mode dipole frequencies and the modified smooth term in the fitting function, the inferred acoustic radius of the glitch is less biased than when only radial modes are used. This remains true, when uncertainties on the simulated frequencies are included in the fitting process.


 {The results from this pilot project  reveal the potential of the proposed approach to use the dipolar modes to characterise the helium glitch in red giants, as well as the drawbacks of using the radial modes alone to that end. The success of this approach depends on the ability to recover the pure acoustic dipole frequencies from the fit to the period spacings that, in turn, is likely to deteriorate as the number of mixed modes per radial order decreases with decreasing stellar luminosity \citep{jiang20}. Future work will aim at understanding the extent of applicability of this method both to model data, by exploring a grid of red-giant models with a range of masses, luminosities and metallicities, and to real data.
}

\section*{Acknowledgements}
 GD acknowledges the support of the University of Paris-Saclay through the internship grant IDEX (grant agreement n$^{\rm o}$4207). This work was supported by FCT - Fundação para a Ciência e a Tecnologia through national funds (PTDC/FIS-AST/30389/2017, UIDB/04434/2020 and UIDP/04434/2020) and by FEDER - Fundo Europeu de Desenvolvimento Regional through COMPETE2020 - Programa Operacional Competitividade e Internacionalização(POCI-01-0145-FEDER-030389). MSC is supported by FCT through a contract (CEECIND/02619/2017).

\section*{Data availability}
The data underlying this article will be shared on reasonable request to the corresponding author.




\bibliographystyle{mnras}
\bibliography{solar-like_v0} 


\end{document}